\documentclass[oldversion]{aa}
\usepackage{graphicx,color}
\usepackage[varg]{txfonts}
\usepackage{aalongtable}
\begin{document}
   \title{Non-thermal radio emission from O-type stars}
   \subtitle{III. Is Cyg OB2 No. 9 a wind-colliding binary?}

   \author{S. Van Loo\inst{1}, R. Blomme\inst{2}, S. M. Dougherty\inst{3} 
	\and M. C. Runacres\inst{4}}
   \authorrunning{Van Loo et al.}

   \offprints{S. Van Loo\\ \email{svenvl@ast.leeds.ac.uk}}

   \institute{School of Physics and Astronomy, University of Leeds, 
	Leeds LS2 9JT, UK \and Royal Observatory of Belgium, Ringlaan 3, 1180
	Brussel, Belgium \and National Research Council of Canada, Herzberg
	Institute for Astrophysics, Dominion Radio Astrophysical Observatory,
	PO Box 248, Penticton, British Columbia V2A 6J9, Canada \and
	Erasmus University College, Nijverheidskaai 170, 1070 Brussel, Belgium}

   \date{Received date; accepted date}

   \abstract{The star Cyg OB2 No. 9 is a well-known non-thermal radio emitter. 
             Recent theoretical work suggests that all such O-stars should 
	     be in a binary or a multiple system. However, there is no 
	     spectroscopic evidence of a binary component. 
             Re-analysis of radio observations from the VLA
	     of this system over 25 years has revealed that the non-thermal 
	     emission varies with a period of $2.35\pm0.02$ yr. This is 
	     interpreted as a strong suggestion of a binary system, with 
	     the non-thermal emission arising in a wind-collision region. 
             We derived some preliminary orbital parameters for this putative 
	     binary and
             revised the mass-loss rate of the primary star downward from 
             previous estimates. 	     
    }

   \maketitle

\section{Introduction}
The Cygnus OB2 association has been the focus of many massive-star 
studies. Observations show that it contains more than 2600 OB stars, 
of which $\sim$ 120 are O stars (Kn\"odlseder \cite{K00}). It also 
harbours some of the hottest and most luminous stars known in our 
Galaxy. Several of these stars exhibit significant variations in their 
X-ray and/or radio continuum emission.  

One of these stars is \object{Cyg OB2 No.~9} (VI~Cyg~9, Schulte~9, MT~431),
which is classified as an O5If star (Massey \& Thompson \cite{MT91}). 
It was one of the first O stars to be identified as a non-thermal 
radio source (Abbott et al.~\cite{ABC84}). It shows the classical 
properties of non-thermal radio emission, such as a high brightness 
temperature (Phillips \& Titus \cite{PT90}) and, also, a radio spectral 
index\footnote{The radio spectral index $\alpha$ describes the 
power-law behaviour of the flux: 
$F_\nu \propto \nu^\alpha \propto \lambda^{-\alpha}.$ For thermal 
emission $\alpha \approx$ +0.6.} with a characteristic negative value.

The non-thermal emission comes from relativistic electrons, believed 
to be accelerated in shocks. Although shocks are ubiquitous in O-star 
winds due to the instability of the radiative driving mechanism (e.g. 
Owocki \& Rybicki~\cite{OR84}), theoretical studies show that this 
embedded-shock model 
cannot be the source of the non-thermal emission, and that the
accelerating shocks arise where the two stellar winds collide (Van Loo
et al.~\cite{VRB06}). This conclusion necessarily implies that all
non-thermal emitting O stars are in a binary system. A similar correlation 
between non-thermal radio emission and binarity is already well established for 
WR stars (Dougherty \& Williams \cite{DW00}) which are the evolutionary 
descendants of O stars. 

At present, there is no spectroscopic evidence that Cyg OB2 No.~9 is a 
binary. De Becker (\cite{PhDM}) and Kiminki et al.~(\cite{K06}) measured 
the radial velocity on a number of occasions. Neither study found 
significant variation on time-scales of a few months to a year. In the 
absence of spectroscopic evidence of binarity, we need to look to
other indicators. Possible techniques include high-spatial resolution 
radio observations that reveal the characteristic shape of a wind-collision 
region 
as observed in \object{WR~140} (Dougherty et al.~\cite{DBC05})
and \object{WR~147} (Williams et al.~\cite{Wetal90}), or searching for 
radio variability
consistent with 
the orbital motion of a binary (e.g. WR140: Williams et al.~\cite{Wetal90}). 

\begin{figure}
\resizebox{\hsize}{!}{\includegraphics{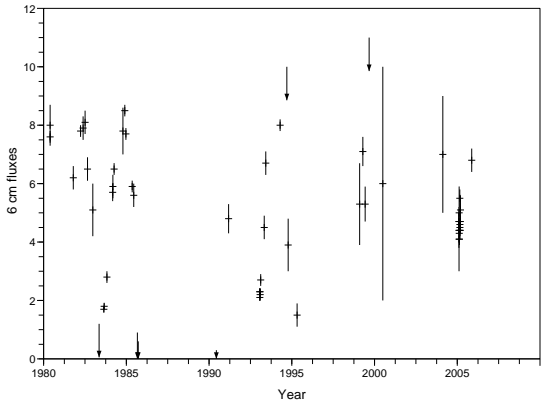}}
\caption{The observed 6-cm fluxes of Cyg~OB2 No.~9 as a function of time.}
\label{fig:temporal}
\end{figure}

In this paper we demonstrate that the radio emission in Cyg OB2 No. 9
has a clearly periodic behaviour. Radio data obtained with the Very Large
Array (VLA) since 1980 were re-reduced, revealing a 2.35-yr period. It
is argued that this is evidence of a binary system, with the non-thermal 
radio emission arising in a wind-collision region.

\section{Data and radio light curve}\label{sect:data}
We collected archive data for Cyg OB~2 No.~9 from the
NRAO Very Large
Array\footnote{The National Radio Astronomy
             Observatory is a facility of the National Science Foundation
             operated under cooperative agreement by Associated Universities,
             Inc.}
archive. Many of these observations have not been published previously.
To avoid introducing systematic effects in the data reduction, we 
re-reduced all observations consistently. Details of the reduction are 
given in Appendix A. The measured fluxes and their error bars (or upper 
limits) are listed in Table A.1.

Using the observations listed in Appendix~\ref{appendix data reduction} 
we can look for variability at radio wavelengths. In 
Fig.~\ref{fig:temporal} we see that the 6~cm fluxes of Cyg OB2 No.~9 
are highly variable with maxima found at e.g. 24 May 1980 and 3 July 1982, 
and minima at e.g. 9 May 1983 and 19 September 1985. These extrema give a 
rough estimate of a period, that is {\it if} any period is present, of the 
order of 2 yr.  However, due to the problem of aliasing, we cannot exclude 
any divisor of 2 yr.

To find periodicity in the data, we apply the string-length technique 
described by Dworetsky (\cite{D82}) to the different wavelength 
observations. The period is found by minimising the distance between the 
flux observations in an orbital phase 
diagram, i.e. the string length. The string-length 
method is particularly useful for a small number of observations with a 
large separation in time. Furthermore it does not require a specific shape 
for the light curve. However, this method does not work when multiple 
periods are present in the data (e.g. Bourguignon et al. \cite{BCB07}).  
Also, this technique neglects the effect of errors on the observation. We, 
therefore, modified the method by replacing the flux difference , i.e. 
$(f_1 - f_2)^2$, by the mean flux-difference, or 
$(f_1 - f_2)^2 + (\sigma_1^2 + \sigma_2^2)$ (where $\sigma$ is the error 
on the flux) when calculating the distance between the observations. 

\begin{figure}
\resizebox{\hsize}{!}{\includegraphics{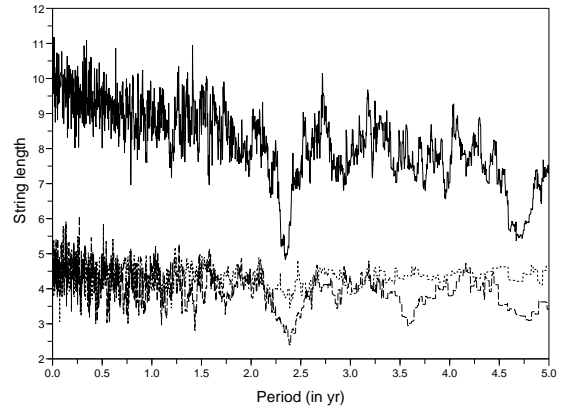}}
\caption{The string length as a function of period for different 
	wavelength observations of Cyg OB2 No. 9:  6 cm (solid), 3.6 cm 
	(dashed), 2 cm (dotted). We do not plot the 20 cm fluxes as it is 
	similar to the 2~cm line. The minimum representing the most-likely 
	period is found near 2.35 yr.}
\label{fig:string}
\end{figure}

\begin{figure*}
\resizebox{\hsize}{!}{\includegraphics{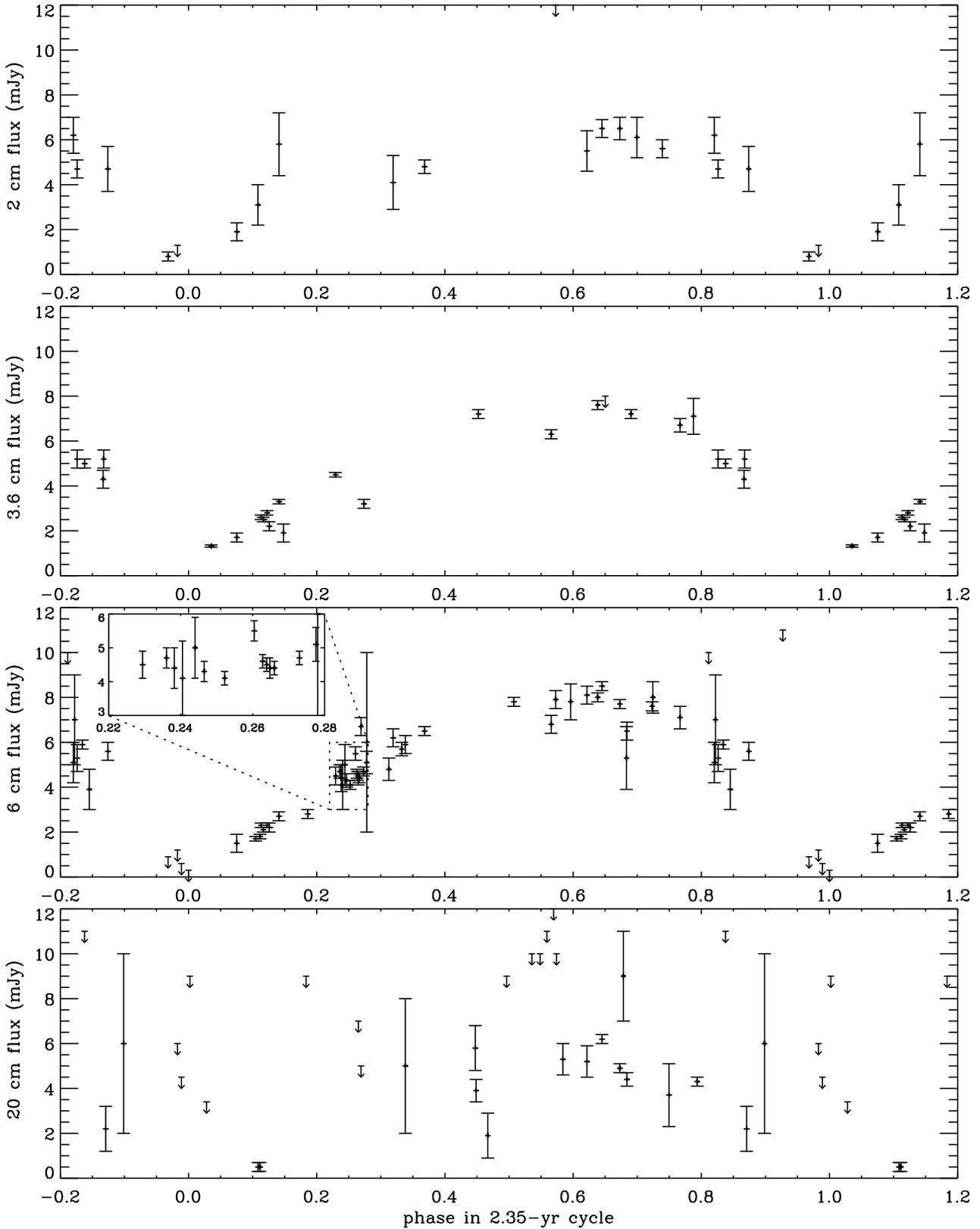}}
\caption{The 2, 3.6, 6 and 20~cm radio fluxes of Cyg OB~2 No.~9,
	folded with a 2.35-yr period. Phase 0.0 has been arbitrarily set 
	at the lowest 6~cm upper limit (JD 2443762.18). }
\label{figure fluxes}
\end{figure*}

We find a minimum in the string length between 2.3 and 2.4 yr in both the 
3.6 and 6~cm fluxes (see Fig.~\ref{fig:string}). However, there is no 
periodicity found in the 2 and 20~cm fluxes.  This is not surprising as 
the 2~cm fluxes have a high degree of noise making a periodicity search 
difficult.  At 20~cm, we do not find a period because most 
observations have a similar flux level or a high upper limit (see 
Fig.~\ref{figure fluxes}). Using the method described by Fernie~(\cite{F89}), 
we can estimate the true period and its uncertainty. We find 
$P = 2.34 \pm 0.03$~yr for the 6~cm fluxes and $P = 2.37 \pm 0.03$~yr 
for the 3.6~cm fluxes. A weighted mean and variance then gives 
$P = 2.355 \pm 0.015$~yr. This fits well with our crude first estimate 
and is confirmed by other period-finding methods such as a discrete 
Fourier transform (Scargle~\cite{S82}). 

To assess the goodness of the period, Fig.~\ref{figure fluxes} shows the 
2, 3.6, 6 and 20~cm radio fluxes of Cyg OB~2 No.~9 folded with the 
2.35-yr period. Phase 0.0 has been arbitrarily set at the lowest 6~cm 
upper limit (JD 2443762.18). The consistency in the light curve shows that 
these reduction techniques do not introduce systematic effects, though 
they may be responsible for a few minor effects. There are some low fluxes 
around phase 0.68 at 6~cm and a somewhat higher flux at phase 0.27 that
we attribute to uncertainty in the absolute flux calibration (see notes
in Table~\ref{table radio data}).

The large number of 6~cm observations around phase 0.23--0.28 (see inset 
to Fig.~\ref{figure fluxes}) could in principle be useful to search for
variations on shorter timescales. There does not seem to be significant 
variability however. A linear regression for this subset shows that, except
for one observation which has a deviation of 3$\sigma$, all observations 
can be fitted. Furthermore, the flux increase is consistent with the 
gradient seen in the main figure.

\section{Discussion}\label{sect:interpretation}
\subsection{Binarity}
The 2.35~yr period derived for the non-thermal radio emission of 
Cyg OB~2 No.~9 is strongly suggestive of a binary system with non-thermal 
radio emission arising in a wind-collision region. The intrinsic synchrotron 
radiation of the wind-collision region changes as a function of the 
binary separation, while free-free absorption removes some of the intrinsic 
emission, depending on the position in the wind (i.e. optical depth along 
the line-of-sight to the observer). Other mechanisms such as inverse-Compton 
cooling and the Razin effect are also dependent on the binary 
separation (e.g. Pittard et al.~\cite{Petal06}). 
These mechanisms therefore produce a radio light curve that 
is modulated by the binary orbit. For WR~140, the best studied example of 
a colliding-wind binary, the radio light curve shows good repeatability 
from one orbital period to another (White \& Becker \cite{WB95}).

The binarity of Cyg OB2 No.~9 could in principle show up in other 
observational diagnostics. We review a number of them below.

\paragraph{Optical observations}
The semi-major axis, $a$, of the binary can be estimated from
\begin{equation}
        (a[AU])^3 = M[M_{\sun}] (P[{\rm yr}])^2,
\label{eq:a3MP2}
\end{equation}
where $M$ is the mass of the binary system. A lower limit on the mass 
of Cyg~OB2 No.~9 can be estimated on the basis of the Eddington limit 
\[
        \frac{\kappa_e L_*}{4 \pi G M_* c} < 1.
\]
Using  $\log(L_*/L_{bol}) = 6.24$ (Herrero et al.~\cite{Hetal99})
we find $M_* > 40 M_{\sun}$. For hot, massive stars, however, the 
Eddington parameter is often within a factor of two below unity, so 
that the stellar mass is more likely to be $M_* > 80 M_{\sun}$. This 
agrees well with observational estimates of the evolutionary mass, 
i.e.  $\sim 100 M_{\sun}$ (Herrero et al.~\cite{Hetal99}).  Although 
we cannot say anything about the actual mass of the binary, it is 
unlikely to exceed $\approx 140 M_{\sun}$, the mass of the most 
massive hot-star binary known, \object{WR20a} (Rauw et al.~\cite{Rauw04}). 
For a binary mass between 80 and 140 $M_{\sun}$, we find a semi-major axis of
$7.6~{\rm AU} \la a \la 9.2$~AU from Eq.~(\ref{eq:a3MP2}). 

The separation between the binary components lies within the
$(1 \pm e) a$ range.  This means that, even for large eccentricities, 
the maximum separation is $\la $ 18~AU. With a distance to the Cyg OB2 
association of 1.82~kpc (Bica et al.~\cite{Betal03}), the components 
have a maximum separation on the plane of only 0.01 arcsec. A Hubble FGS 
(Project: FGS 10612; PI: D. Gies \footnote{URL: 
http://www.stsci.edu/observing/phase2-public/10612.pro}) observation of 
Cyg OB No.~9 in April 2006 looked for a binary companion with luminosity 
difference less than 4 magnitudes within a 0.01 - 1 arcsec region around 
the primary. Given 
the estimated separation, it is unlikely that a companion 
can be detected using FGS.

\paragraph{Radial velocity measurements} Two independent studies of 
radial velocity variations by De Becker (\cite{PhDM}) and Kiminki et 
al.~(\cite{K06}) did not find any significant changes in radial velocity.
However, by combining the measurements of these studies, we find that a 
constant radial velocity can be rejected on a 97~\% confidence level.
Thus, Cyg~OB2 No.~9 is a possible binary by the criterion of Kiminki et 
al.~(\cite{K06}).  Furthermore, re-examination
of spectroscopic archive data shows a possible double-lined spectrum
(Kiminki, priv. comm.). Additional spectroscopic observations are needed 
to confirm the preliminary spectroscopic evidence of binarity. 

With the derived constraints on the binary separation and the period,
the expected amplitude of the radial velocity variations can be calculated.
The observed variation is likely to be due to the motion of the primary.
For a Keplerian orbit, the radial velocity amplitude is given by
\begin{equation}\label{eq:RV} 
       K_{p} = \frac{2\pi a_{p}}{P} \frac{\sin i}{\sqrt{1-e^2}}
        = \frac{2\pi a}{P}\frac{M_s}{M_p+M_s} \frac{\sin i}{\sqrt{1-e^2}},
\end{equation}
where $a_p$ and $a$ are the semi-major axis of the primary star and the 
binary system respectively, $M_p$ and $M_s$ the mass of the primary and 
secondary star respectively,
$i$ the inclination angle and $e$ the eccentricity. For a  
$100 + 40~M_{\sun}$ binary, $a_p = 0.28 a$ which gives 
$K_p = 32 \sin i/\sqrt{1-e^2}~{\rm km~s^{-1}}$. For a  
$70 + 70~M_{\sun}$ binary, one finds 
$K_p = 55 \sin i/\sqrt{1-e^2}~{\rm km~s^{-1}}$. This can then be
compared to the observed radial velocity amplitude. By combining both data 
sets of De Becker (\cite{PhDM}) and Kiminki et al.~(\cite{K06}), we 
find a  marginally significant variation  with 
a radial velocity amplitude $K_{p}$ between 27 and $69~{\rm km~s^{-1}}$.

\begin{figure}
\resizebox{\hsize}{!}{\includegraphics{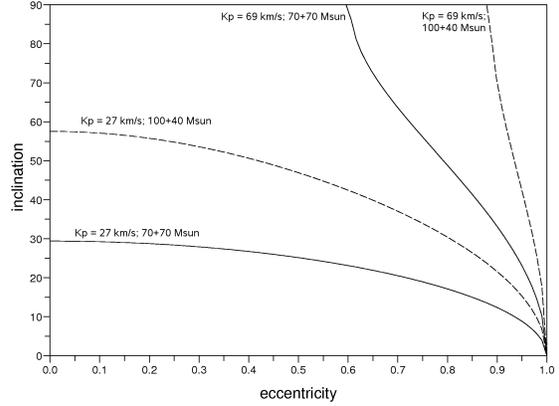}}
\caption{Contour lines for $\sin i/\sqrt{1-e^2}$, where  
	the solid line corresponds to a $70+70~M_{\sun}$ binary
	and the dashed to a $100 + 40~M_{\sun}$ binary. 
        The permitted range of eccentricity and inclination lies within 
        the contour lines. }
\label{fg:contour}
\end{figure}

We find that this comparison allows a first estimate of the orbital
parameters of the system, through the constraint on the
function $\sin i/\sqrt{1-e^2}$. Figure~\ref{fg:contour} shows the 
permitted values of the eccentricity and the inclination
for different binary mass ratios. Due to the broad range of possible 
radial velocity amplitudes, we only find a weak constraint on the 
inclination and eccentricity. However, the derived value for the 
radial velocity variation is subject to considerable uncertainty: 
the sampling of the orbit is sparse, which makes it likely that the maximum 
variation was missed. More recent observations 
suggest that the radial velocity amplitude tends to the upper
value of our adopted range (Naz\'e, pers. comm.). This favours 
a high eccentricity orbit ($e > 0.6$) viewed at a substantial 
inclination, e.g. $i > 60^\circ$. 
 
\paragraph{X-ray emission} 
A non-thermal radio emitting binary should produce both thermal and 
non-thermal X-rays. The non-thermal component is produced by 
inverse-Compton cooling of relativistic electrons in the strong stellar 
ultraviolet 
radiation field. The thermal component is produced by the hot gas 
($\sim 10^6$~K)
produced in the wind collision. For a long period binary such as  
Cyg OB2 No.~9, this thermal component is hard, and dominates the 
non-thermal contribution in the 0.5 -- 10 keV range (De Becker \cite{PhDM}).
Depending on orbital parameters, some modulation of the X-ray emission with 
the orbital period can be expected. However, the X-ray emission of 
Cyg OB2 No. 9 remains relatively stable for $\sim$ 15 years (Waldron et 
al.~\cite{W98}). 
However, it should be noted that the X-ray observations 
are sparsely distributed throughout the orbital phase space, and variations 
due to orbital changes might have been  missed. 

By fitting the X-ray data with a two-temperature model, Rauw et 
al.~(\cite{Retal05}) find that the higher temperature of the two, i.e. 
12$\times 10^6$~K, agrees well with the temperatures expected in a 
wind-collision binary (Stevens et al.~\cite{SBP92}). Interestingly, 
the observed X-ray emission from No. 9 is smaller than the canonical 
$L_X/L_{bol} = 10^{-7}$ (Waldron et al.~\cite{W98}). Although some scatter 
around this value can be expected, binaries are often much brighter in 
X-rays than single stars (Pollock~\cite{P87}). In this regard, Cyg OB2 
No.~9 does not have a typical X-ray luminosity for a binary.

\paragraph{Emission line variability}
Another possible indicator of colliding winds is the shape and variability
of the emission lines. In the colliding-wind binary \object{Cyg OB2 No. 5},
the He II $\lambda$4686 line shows orbital phase variations, but with a
phase offset compared to the photospheric lines. This shows that at least
part of this line is formed in the colliding-wind region 
(Rauw et al.~\cite{Retal99}). Similar effects could be present in 
Cyg OB2 No. 9.

\paragraph{High-spatial resolution radio observations}
Recent high-resolution radio observations of Cyg~OB2 No.~9 with the VLBA
reveal a bow-shaped emission region (Dougherty et al., in prep). This is
consistent with emission arising in a wind-collision region, and is very
reminiscent of the emission distribution observed in proto-typical
colliding-wind systems like WR~140 (Dougherty et al.~\cite{DBC05}). 
This is the strongest direct
evidence to date that the non-thermal radio emission in Cyg~OB2  No.~9 
arises due to a colliding-wind binary.

\subsection{Spectral index}
\begin{figure}
\resizebox{\hsize}{!}{\includegraphics{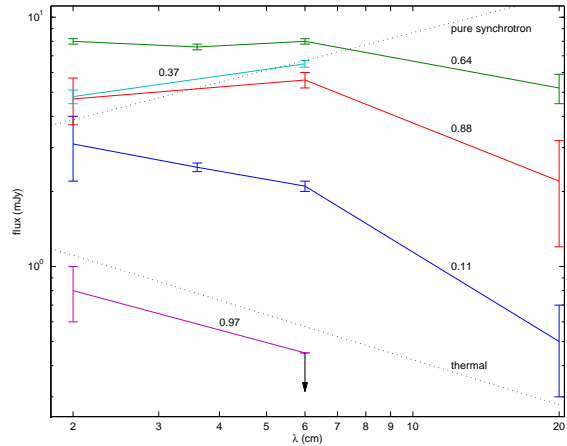}}
\caption{The radio spectrum at chosen orbital phases. A thermal radio 
	spectrum and a pure synchrotron spectrum are plotted at arbitrary 
	flux levels, for the sake of reference. The numbers refer to 
	the orbital phase of the observation.}
\label{fg:specindices}
\end{figure}

Fig.~\ref{fg:specindices} shows the fluxes at chosen orbital phases,
as a function of the available wavelengths. Globally, the radio spectrum 
becomes shallower as the orbit progresses from radio minimum (phase 
$\varphi \approx 0$) to radio maximum ($\varphi \approx 0.65$)\footnote{Recall 
that a positive slope on the figure corresponds to a 
negative spectral index, and vice versa.} Near radio minimum
($\varphi = 0.97)$, the spectral index between 2 and 6~cm is 
$\alpha_{26} > 0.26$, suggesting an optically-thick spectrum with
substantial attenuation, most likely due to free-free absorption. One 
may surmise that the minimum occurs near periastron, where the wind-collision 
region is buried in the stellar wind,
though we readily admit
this is not the only possible configuration for a minimum. The shape of the 
spectrum near radio minimum suggests thermal emission, but does not exclude
a modest non-thermal contribution. Between phase 0.1 and 0.3 
the spectral index $\alpha_{26}$ switches from positive to negative.
The spectral index between 6 and 20~cm is always negative. At radio 
maximum the spectrum has a turnover from negative to positive spectral 
index as one goes from short to long wavelengths. This is consistent 
with the attenuation expected from free-free absorption, which is greater 
at larger wavelengths. The spectral index at maximum is
$\alpha_{26} = -0.24 \pm 0.09$, which differs from the -0.5 index expected
for synchrotron emission arising from strong-shock acceleration. This 
suggests that, even at radio maximum, the observed flux is a combination 
of optically thick and optically thin emission. 

\subsection{Mass-loss rate}
We can derive an upper limit for the mass-loss rate from the radio minimum, 
where we can expect the contribution of non-thermal emission to be smallest. 
Adopting the stellar parameters given in Van Loo et al. (\cite{VanLoo+2004})
and taking the $0.3$ mJy upper limit at 6 cm (11 June 1990), we find 
$\dot{M} < 0.9\times 10^{-5}~{\rm M_{\sun}~yr^{-1}}$. The 0.8 mJy detection 
at 2 cm gives $\dot{M} < 1.1\times 10^{-5}~{\rm M_{\sun}~yr^{-1}}$. These 
values are upper limits, as the observed flux at radio minimum may still 
have a non-thermal contribution. They are lower than the
$\dot{M}$(radio) =  1.9 $\times 10 ^{-5}~{\rm M_{\sun}~yr^{-1}}$ given by
Abbott et al.~(\cite{ABC84}), partly because of the lower observed flux,
and partly because of improved stellar and wind parameters.

The mass-loss rate originally derived from H$\alpha$ is 
1.77$\times 10^{-5}~{\rm M_{\sun}~yr^{-1}}$ (Leitherer~\cite{L82}),
which yields 1.47$\times 10^{-5}~{\rm M_{\sun}~yr^{-1}}$ with the parameters
given in Van Loo et al. (\cite{VanLoo+2004}). A recent study by Puls 
et al.~(\cite{P06}) shows that, for O star supergiants with H$\alpha$ in 
emission, $\dot{M}$(radio) $\approx 0.4-0.5~\dot{M}$(H$\alpha$). This is 
believed to be caused by a radial decrease in the amount of clumping, as 
is also predicted theoretically (Runacres \& Owocki \cite{RO05}). 
We find for Cyg OB2 No. 9 that $\dot{M}$(radio) $< 0.61 \dot{M}$(H$\alpha$), 
which is roughly consistent with the results of Puls et al.~(\cite{P06}).

\subsection{Energy budget of the non-thermal radio emission}
The radio synchrotron emission is maintained by the kinetic power of the
stellar winds, but only part of that is dissipated through the
wind-collision shocks, and only a fraction of that power is dissipated in
relativistic electrons. In addition, radio synchrotron emission accounts for
a small fraction of the total non-thermal energy, with the bulk of that
energy radiated at high energies.

There is considerable uncertainty how much of the wind-collision power is
converted into radio synchrotron luminosity. Following Chen \& White 
(\cite{CW94}) and Pittard \& Dougherty (\cite{PD06}) we suggest the following 
order-of-magnitude estimate of
\[ 
L_{\rm syn} \approx 10^{-4}L_{\rm NTe} \approx 5\times 10^{-7} L_{\rm WC},
\]
where $L_{\rm NTe}$ is the power in non-thermal electrons and $L_{\rm WC}$
is the kinetic power of the winds that enter the wind-collision region.

The kinetic power 
$L_{\rm WC}$ depends strongly on the momentum ratio of the winds. For equal
winds, the kinetic power in the wind-collision region is roughly half of the
total kinetic power of the winds, whereas for a system with a much smaller
momentum ratio e.g WR\thinspace140, the kinetic power in
the winds is only 1\% of the total wind power due to the small solid angle
presented by the wind-collision region to the dominant wind (Pittard \&
Dougherty \cite{PD06}). For Cyg~OB2 No.~9, using the stellar wind parameters, we
estimate that  $L_{\rm syn}\approx 10^{31}$~erg~s$^{-1}$ if the wind
momenta are assumed to be equal, and $L_{\rm sync} \approx 10^{29}$~erg~
s$^{-1}$ if the companion has a much weaker wind.
The observed radio synchrotron luminosity $ \sim 10^{30}$~erg~ s$^{-1}$,
which is likely a lower limit on the intrinsic synchrotron emission due to
attenuation of emission by absorption and/or cooling. In support of these
numbers, the model of Eichler \& Usov (\cite{EU93}) gives a similar value. Given
all the uncertainties in the above discussion, these numbers are all closely
compatible, supporting a wind-collision origin for the non-thermal emission.

\section{Conclusions}\label{sect:conclusions}
The most important result of this paper is that 
a period of 2.35 yr was found in the radio emission from
Cyg~OB2 No.~9. This is 
highly suggestive of a binary system in which the non-thermal emission, 
modulated by the orbit, arises in a wind-collision region. The binary 
status of Cyg OB2 No. 9 can currently not be confirmed by spectroscopic, 
optical or X-ray observations. However, some of the observations 
support the presence of a binary
e.g. the  high temperature derived from the X-ray 
spectrum and the marginally significant radial velocity variations. 
Additional observations are 
required to substantiate the claim of a binary using these techniques. 
However, there is already compelling evidence
of a wind-collision region based on high-resolution radio observations
(Dougherty et al., in prep).

From the radio minimum we derive an upper limit for the mass-loss rate 
of $0.9\times 10^{-5}~{\rm M_{\sun}~yr^{-1}}$. This is substantially 
lower than previous estimates. As the radio formation region is likely 
to be less affected by clumping than the formation region of H$\alpha$, 
we suggest our upper limit is the most reliable estimate to date.

\begin{acknowledgement}
We thank Daniel Kiminki, Michael De~Becker and Yael Naz\'e for 
useful discussions and 
providing valuable radial velocity data. SVL gratefully acknowledges 
STFC for the financial support. We thank Joan Vandekerckhove for his help 
with the reduction of the VLA data. We are grateful to the original 
observers of the data we used from the VLA archive.  This research has 
made use of the SIMBAD database, operated at CDS, Strasbourg, France, 
NASA's Astrophysics Data System Abstract Service and data from the 
University of Michigan Radio Astronomy Observatory which is supported by funds 
from the University of Michigan. MCR thanks Tom Hartquist for his hospitality
and kind support during a stay at the University of Leeds to prepare this 
paper.
\end{acknowledgement}

\Online
\appendix
\section{Data reduction}
\label{appendix data reduction}
The data reduction was accomplished using the Astronomical Image
Processing System (AIPS), developed by the NRAO, following the
standard procedure of antenna gain calibration, absolute flux
calibration, imaging, and deconvolution. These have been described in
detail in Paper~I.

For some of the observing programmes that we examined, no flux
calibrator was observed and we used calibration observations from
observing programmes at closely the same time. For programmes VP51 and
AL372 at 20cm no such observations were available and we resorted to
the value listed in the VLA calibrator manual\footnote{URL:
http://www.vla.nrao.edu/astro/calib/manual} making the fluxes from
these observations less reliable. As a consistency check on the
derived flux scale, we compared fluxes for the most commonly used
phase-reference source, J2007+404, with those obtained at 2, 3.6, and
6~cm at the University of Michigan Radio Astronomy Observatory
(UMRAO). Excellent agreement was found for all observations except
AC116 (1985-01-13) at 2cm. For this specific case, the flux scale used 
was derived from interpolation of the MRAO data.

Antenna gain calibration was established in the standard manner of
self-calibration of the phase-reference source and transferring these
solutions to the target field-of-view. These solutions were then
improved further by one round of phase-only calibration of the target
field-of-view. In some cases, smoothly varying phase solutions were
not found using the additional target self-calibration, and so we
adopted the phase-referenced solutions. These cases where target
self-calibration was applied are noted in Table~\ref{table radio data}.

Images of the field-of-view were derived by filtering out the lowest
spatial frequency data. This eliminates large scale "background"
emission that is particularly prevalent at 20cm in the more compact 'C'
and 'D' configuration observations. This should have no effect on the
flux estimates of Cyg OB2 No. 9 since this source is unresolved by the
VLA in the highest resolution 'A' configuration at all wavelengths examined.

After image deconvolution using {\sc AIPS} task {\sc CLEAN}, the flux
of Cyg OB2 No. 9 was established by fitting elliptical Gaussians to
the source image. The resulting fluxes are listed in 
Table~\ref{table radio data}. Correction for primary beam attenuation has 
been made for observations where Cyg OB2 No. 9 is not at the field centre. 
Some care has been taken to assess the uncertainty in the derived fluxes. The
uncertainties given in Table\ref{table radio data} include not only the 
root-mean-square (RMS) noise in the map, but also an estimate of the 
uncertainty in the absolute flux calibration process (5\% at 0.7 and 2cm, 
and 2\% at 3.6, 6, 20 and 90~cm) and an estimate of systematic errors.  
The systematic errors were evaluated using a jackknife technique, described 
in Paper I. In those observations where Cyg OB2 No. 9 was not detected, an
upper limit is quoted that is 3 times the uncertainty as derived above.

A comparison between the fluxes derived here and those in the
literature show good agreement overall, with sometimes slightly higher
values reported here, perhaps due to the use of phase self-calibration
on the target field. There are a few notable exceptions to this
agreement. In the AC42 data at 2 and 6cm, there is weak
($\sim3\sigma$) source at the location of Cyg OB2 No. 9, but other
stronger "sources" nearby do not repeat in other 2 and 6cm
observations. Hence we assign a $3\sigma$ upper limit (contrary to
Bieging et al.~\cite{BAC89}).In the BIGN data, we find a $4\sigma$ detection,
where Bieging et al. provide a $3\sigma$ upper limit. We do not detect
Cyg OB2 No. 9 in VP91 at 20cm as reported by Phillips \& Titus
(\cite{PT90}). This is a D-configuration observation with considerable large
scale emission around the position of Cyg OB2 No. 9. It is not possible to
totally eliminate this emission by removing the low spatial frequency
data since it leaves insufficient data for successful imaging, but
examination of 20cm data from higher-resolution configurations shows
that this is background emission unrelated to Cyg OB2 No. 9. We
speculate that Phillips \& Titus may have mis-identified some of this
emission as Cyg OB2 No. 9. In the AR328 0.7cm data, we find a weak
"source" at the position of No. 9, but again there are brighter nearby
sources that do not appear at other frequencies. Hence we have adopted
an upper limit contrary the detection reported by Contreras et
al. (\cite{Contreras+al96}). On the other hand, we do detect No. 9 at both 
3.6 and 6cm from AR328.

Lastly, we note a number of reporting errors. Bieging et al. (\cite{BAC89})
list incorrectly some observations dated 1982 Feb 9. Also, Waldron et
al (\cite{W98}) give their 1991 observation as being at 6cm, rather than 3.6cm,
and their 1992 data as taken on Jun 24, rather than Jan 24.

{\footnotesize
\begin{longtable}{lllrrcrrrlcl}
\caption{\label{table radio data}Reduction of the VLA data. 
Column (1) gives the programme name, (2) the date of the observation,
(3) the source on which the observation was centred,
(4) the phase calibrator name (J2000 coordinates),
(5) the phase calibrator flux (Jy) and 
(6) distance to the observation centre (degrees),
(7) the integration time (in minutes) on the source,
(8) the number of antennas that gave a usable signal,
(9) the configuration the VLA was in at the time of the observation,
(10) the measured flux (in mJy) and
(11) refers to the notes.
Many of the VLA observations were made in two sidebands, each of which has a 
bandwidth of 50 MHz; the exceptions are noted in column (11).
Upper limits are 3 $\times$ the RMS.
Numbers between brackets in the notes column give references for those
observations that have already been published in the literature.
}\\
\hline\hline
& \multicolumn{1}{c}{(1)} & \multicolumn{1}{c}{(2)} & \multicolumn{1}{c}{(3)} &
\multicolumn{1}{c}{(4)} & \multicolumn{1}{c}{(5)} & \multicolumn{1}{c}{(6)} &
\multicolumn{1}{c}{(7)} & \multicolumn{1}{c}{(8)} & \multicolumn{1}{c}{(9)} &
\multicolumn{1}{c}{(10)} & \multicolumn{1}{l}{(11)} \\
& \multicolumn{1}{c}{progr.} & \multicolumn{1}{c}{date} &
\multicolumn{1}{c}{ctr.} &
\multicolumn{3}{c}{phase calibrator} &
\multicolumn{1}{c}{int.} &
\multicolumn{1}{c}{no.} &
\multicolumn{1}{c}{config.} &
\multicolumn{1}{c}{flux} &
\multicolumn{1}{l}{notes}\\
\cline{5-7}
&  &  & &
\multicolumn{1}{c}{name} &
\multicolumn{1}{c}{flux} &
\multicolumn{1}{c}{dist.} &
\multicolumn{1}{c}{time} &
\multicolumn{1}{c}{ants.} & &
\multicolumn{1}{c}{(mJy)} & \\
\hline
\endfirsthead
\caption{continued.}\\
\hline\hline
& \multicolumn{1}{c}{(1)} & \multicolumn{1}{c}{(2)} & \multicolumn{1}{c}{(3)} &
\multicolumn{1}{c}{(4)} & \multicolumn{1}{c}{(5)} & \multicolumn{1}{c}{(6)} &
\multicolumn{1}{c}{(7)} & \multicolumn{1}{c}{(8)} & \multicolumn{1}{c}{(9)} &
\multicolumn{1}{c}{(10)} & \multicolumn{1}{l}{(11)} \\
& \multicolumn{1}{c}{progr.} & \multicolumn{1}{c}{date} &
\multicolumn{1}{c}{ctr.} &
\multicolumn{3}{c}{phase calibrator} &
\multicolumn{1}{c}{int.} &
\multicolumn{1}{c}{no.} &
\multicolumn{1}{c}{config.} &
\multicolumn{1}{c}{flux} &
\multicolumn{1}{l}{notes}\\
\cline{5-7}
&  &  & &
\multicolumn{1}{c}{name} &
\multicolumn{1}{c}{flux} &
\multicolumn{1}{c}{dist.} &
\multicolumn{1}{c}{time} &
\multicolumn{1}{c}{ants.} & &
\multicolumn{1}{c}{(mJy)} & \\
\hline
\endhead
\hline
\endfoot
\multicolumn{12}{l}{
 \begin{tabular}{lllllllllllll}
 \multicolumn{2}{l}{Notes:} \\
\multicolumn{5}{l}{Column (3): source on which the observation was centred
is mostly indicated by Schulte (\cite{Schulte56}) Cyg OB~2 numbers.
Others indicators are:} \\
\multicolumn{5}{l}{`98' is in between No. 8A and No. 9; `+' is listed in the observing log as
2031+411, `X3' as Cyg X-3, `XA' as Cyg X-3A, `23' as C23,} \\
\multicolumn{5}{l}{`80' as F80XF, `21' as 2032+41, `91' as 20315+41291, `32' 
as 20330+41132, `20' as 2033+40, `J2' as 2033+4118\_2 and `SE' as TEVSE.} \\
1x50  & (or a similar code): bandwidth different from standard 2x50 MHz \\
C & flux calibration based on MRAO data (see text) \\
I & stripes are visible on this image, complicating the flux measurement  \\
PH & flux calibration on phase calibrator \\
S & selfcalibration applied \\
X   & flux calibration is very uncertain; check on flux of phase calibrator suggests factor of 2 \\
(1) & White \& Becker (\cite{White+Becker83}) \\ 
(2) & Bieging et al.~(\cite{BAC89}) \\ 
(3) & Phillips \& Titus (\cite{PT90}) \\ 
(4) & Contreras et al.~(\cite{Contreras+al96}) \\ 
(5) & Waldron et al.~(\cite{W98}) \\ 
\end{tabular}
}
\endlastfoot
\multicolumn{5}{l}{{\bf 0.7 cm}}\\*
& AR277  & 1994-04-17 & 9 & 2007+404 & 1.18  $\pm$ 0.03  &  4.9 &  32 &  9 &  A  &        $<$  6.    & (4) \\
& AR328  & 1995-04-27 & 9 & 2007+404 & 1.27  $\pm$ 0.01  &  4.9 &  23 & 10 &  D  &        $<$  4.    & (4) \\
\multicolumn{5}{l}{{\bf 1.3 cm}}\\*
& AW515  & 1999-06-08 & 9 & 2007+404 & 1.46  $\pm$ 0.02  &  4.9 &  40 & 19 & AD  & 3.5   $\pm$ 0.6   & \\
\multicolumn{5}{l}{{\bf 2 cm}}\\*
& BIGN   & 1981-10-16 & 9 & 2007+404 & 4.8   $\pm$ 0.6   &  4.9 &  29 & 27 &  C  & 4.1   $\pm$ 1.2   & 1x50,(2) \\
& BECK   & 1982-05-22 & 9 & 2007+404 & 6.1   $\pm$ 1.3   &  4.9 &  13 & 27 &  A  &        $<$ 12.    & 1x50 \\
& BECK   & 1982-07-03 & 9 & 2007+404 & 4.50  $\pm$ 0.08  &  4.9 &   9
& 25 &  A  & 5.5   $\pm$ 0.9   & 1x50,(1) \\
& BECK   & 1982-12-21 & 9 & 2023+318 & 3.18  $\pm$ 0.04  &  9.6 &  14 & 26 &  D  & 6.2   $\pm$ 0.8   & 1x50 \\
& AC42   & 1983-05-09 & 9 & 2007+404 & 3.76  $\pm$ 0.06  &  4.9 &  21 & 27 &  C  &        $<$  1.3   & 1x50,(2) \\
& AB228  & 1983-08-25 & 9 & 2007+404 & 6.9   $\pm$ 0.7   &  4.9 &  19 & 26 &  A  & 3.1   $\pm$ 0.9   & \\
& AA29   & 1984-04-04 & 9 & 2007+404 & 3.53  $\pm$ 0.08  &  4.9 &  22 & 27 &  C  & 4.8   $\pm$ 0.3   & S,(2) \\
& AC116  & 1984-11-27 & 9 & 2007+404 & 3.75  $\pm$ 0.06  &  4.9 &  38 & 25 &  A  & 6.5   $\pm$ 0.4   & (2) \\
& AC116  & 1984-12-21 & 9 & 2007+404 & 3.96  $\pm$ 0.07  &  4.9 &  10 & 27 &  A  & 6.5   $\pm$ 0.5   & S,(2) \\
& AC116  & 1985-01-13 & 9 & 2007+404 & 3.3   $\pm$ 0.2   &  4.9 &  28 & 25 &  A  & 6.1   $\pm$ 0.9   & C,(2) \\
& AC116  & 1985-02-16 & 9 & 2007+404 & 3.45  $\pm$ 0.02  &  4.9 &  19 & 25 &  A  & 5.6   $\pm$ 0.4   & (2)\\
& AA47   & 1985-06-12 & 9 & 2007+404 & 3.2   $\pm$ 0.1   &  4.9 &  15 & 25 &  B  & 4.7   $\pm$ 1.0   & S,(2) \\
& AA47   & 1985-09-01 & 9 & 2007+404 & 3.02  $\pm$ 0.05  &  4.9 &  15 & 27 &  C  & 0.8   $\pm$ 0.2   &  \\
& AB671  & 1993-02-14 & 9 & 2007+404 & 5.8   $\pm$ 0.2   &  4.9 &   6 & 27 & BnA & 5.8   $\pm$ 1.4   &  \\
& AR328  & 1995-04-27 & 9 & 2007+404 & 2.40  $\pm$ 0.02  &  4.9 &  34 & 15 &  D  & 1.9   $\pm$ 0.4   & (4) \\
& AW515  & 1999-06-08 & 9 & 2007+404 & 1.71  $\pm$ 0.01  &  4.9 &  30 & 20 &  AD & 4.7   $\pm$ 0.4   & I \\
\multicolumn{5}{l}{{\bf 3.6 cm}}\\*
& AS397  & 1990-02-16 &22 & 2007+404 & 3.10  $\pm$ 0.04  &  4.9 &  14 & 22 &  A  & 4.3   $\pm$ 0.4   &  \\
& AH395  & 1990-02-17 &22 & 2007+404 & 3.05  $\pm$ 0.01  &  4.9 &  29 & 22 &  A  & 5.2   $\pm$ 0.4   &  \\
& AW288  & 1991-07-04 & 9 & 2025+337 & 2.08  $\pm$ 0.06  &  7.7 &   5 & 26 &  A  & 7.2   $\pm$ 0.2   &  S,(5) \\
& AW304  & 1992-01-24 & 9 & 2007+404 & 2.87  $\pm$ 0.03  &  4.9 &  12 & 24 & CnB & 7.2   $\pm$ 0.2   &  S,(5)  \\
& BP1    & 1992-05-30 & 9 & 2052+365 & 1.87  $\pm$ 0.01  &  6.0 &  99 & 23 & DnC & 5.0   $\pm$ 0.2   &  \\
& AB671  & 1993-01-21 & 9 & 2007+404 & 3.20  $\pm$ 0.06  &  4.9 &  19 & 24 &  A  & 2.6   $\pm$ 0.1   &  S  \\
& AB671  & 1993-01-24 & 9 & 2007+404 & 3.33  $\pm$ 0.07  &  4.9 &  23 & 27 &  A  & 2.5   $\pm$ 0.1   &  S  \\
& AB671  & 1993-01-29 & 9 & 2007+404 & 3.33  $\pm$ 0.02  &  4.9 &  24 & 27 & BnA & 2.8   $\pm$ 0.1   &  S  \\
& AB671  & 1993-02-01 & 9 & 2007+404 & 3.29  $\pm$ 0.03  &  4.9 &   6 & 27 & BnA & 2.2   $\pm$ 0.2   &  \\
& AB671  & 1993-02-14 & 9 & 2007+404 & 3.41  $\pm$ 0.03  &  4.9 &   7 & 27 & BnA & 3.3   $\pm$ 0.1   &  S \\
& AB671  & 1993-02-20 & 9 & 2007+404 & 3.8   $\pm$ 0.1   &  4.9 &  14 & 24 &  B  & 1.9   $\pm$ 0.4   &  \\
& AS483  & 1993-05-01 & 98& 2007+404 & 3.28  $\pm$ 0.01  &  4.9 &  25 & 27 &  B  & 4.5   $\pm$ 0.1   &  S,(5) \\
& AR277  & 1994-04-17 & 9 & 2007+404 & 2.97  $\pm$ 0.01  &  4.9 &  23 & 14 &  A  & 7.6   $\pm$ 0.2   &  S,(4) \\
& AR328  & 1995-04-27 & 9 & 2007+404 & 2.77  $\pm$ 0.01  &  4.9 &  11 & 16 &  D  & 1.7   $\pm$ 0.2   &  (4)\\
& AS644  & 1999-04-18 & 9 & 2015+371 & 2.17  $\pm$ 0.01  &  4.9 &   5 & 27 &  D  & 6.7   $\pm$ 0.3   &  S \\
& AW515  & 1999-06-08 & 9 & 2007+404 & 2.02  $\pm$ 0.01  &  4.9 &  10 & 19 & AD  & 5.2   $\pm$ 0.4   &  S,I \\
& BB116  & 1999-12-04 & 9 & 2007+404 & 2.20  $\pm$ 0.01  &  4.9 & 101 & 27 &  B  & 1.32  $\pm$ 0.05  &  \\
& BB116  & 2000-06-26 & 9 & 2007+404 & 2.40  $\pm$ 0.01  &  4.9 &  20 & 26 & DnC & 3.2   $\pm$ 0.2   &  \\
& BD110  & 2005-11-15 & 9 & 2007+404 & 2.29  $\pm$ 0.02  &  4.9 &  21 & 22 &  D  & 6.3   $\pm$ 0.2   &  S \\
& AB1195 & 2006-01-26 & 8 & 2007+404 & 2.37  $\pm$ 0.05  &  4.9 &  31 & 16 & AD  &         $<$ 8.    &  \\
& AB1210 & 2006-05-24 & 8 & 2007+404 & 2.21  $\pm$ 0.02  &  4.9 &  27 & 21 & BnA & 7.1   $\pm$ 0.8   &  \\
\multicolumn{5}{l}{{\bf 6 cm}}\\*
& CHUR   & 1980-05-23 & 9 & 2007+404 & 4.73  $\pm$ 0.03  &  4.9 &  49 & 22 & AD  & 7.6   $\pm$ 0.2   &  1x50,(2) \\
& CHUR   & 1980-05-24 & 9 & 2007+404 & 4.58  $\pm$ 0.03  &  4.9 &  50 & 23 & AD  & 8.0   $\pm$ 0.7   &  1x50,(2) \\
& BIGN   & 1981-10-16 & 9 & 2007+404 & 4.16  $\pm$ 0.04  &  4.9 &  17 & 27 & C   & 6.2   $\pm$ 0.4   &  1x50,S,(2) \\
& BECK   & 1982-03-27 & 9 & 2007+404 & 4.74  $\pm$ 0.01  &  4.9 & 211 & 27 & A   & 7.8   $\pm$ 0.2   &  1x50,(1) \\
& BECK   & 1982-05-22 & 9 & 2007+404 & 4.59  $\pm$ 0.03  &  4.9 &   5 & 24 & A   & 7.9   $\pm$ 0.4   &  1x50,S \\
& BECK   & 1982-07-03 & 9 & 2007+404 & 4.95  $\pm$ 0.02  &  4.9 &   5 & 26 & A   & 8.1   $\pm$ 0.4   &  1x50,S,(1)  \\
& BIEG   & 1982-08-26 & 9 & 2007+404 & 4.4   $\pm$ 0.1   &  4.9 &   9 & 24 & B   & 6.5   $\pm$ 0.4   &  1x50,(2) \\
& BECK   & 1982-12-21 & 9 & 2023+318 & 2.86  $\pm$ 0.01  &  9.6 &   9 & 24 & D   & 5.1   $\pm$ 0.9   &  1x50,S  \\
& AC42   & 1983-05-09 & 9 & 2007+404 & 4.46  $\pm$ 0.02  &  4.9 &  10 & 26 & C   &         $<$ 1.2   &  1x50,(2) \\
& AC42   & 1983-08-22 & 9 & 2007+404 & 4.85  $\pm$ 0.01  &  4.9 &  14 & 25 & A   & 1.7   $\pm$ 0.1   &  (2) \\
& AB228  & 1983-08-28 & 9 & 2007+404 & 4.90  $\pm$ 0.03  &  4.9 &  18 & 27 & A   & 1.8   $\pm$ 0.1   &  \\
& AB252  & 1983-10-30 & 9 & 2007+404 & 4.84  $\pm$ 0.02  &  4.9 &   9 & 26 & A   & 2.8   $\pm$ 0.2   &  S  \\
& AA28   & 1984-03-04 & 9 & 2007+404 & 4.27  $\pm$ 0.01  &  4.9 &  13 & 27 & CnB & 5.7   $\pm$ 0.3   &  S,(2) \\
& AA28   & 1984-03-09 & 8 & 2007+404 & 4.33  $\pm$ 0.01  &  4.9 &  23 & 26 & CnB & 5.9   $\pm$ 0.4   &   \\
& AA29   & 1984-04-04 & 9 & 2007+404 & 4.21  $\pm$ 0.02  &  4.9 &  11 & 27 & C   & 6.5   $\pm$ 0.2   &  S,(2)  \\
& VM59   & 1984-10-17 & 9 & 2202+422 & 2.72  $\pm$ 0.01  & 16.7 &  13 & 26 & D   & 7.8   $\pm$ 0.8   &  S  \\
& AC116  & 1984-11-27 & 9 & 2007+404 & 4.50  $\pm$ 0.01  &  4.9 &  19 & 25 & A   & 8.5   $\pm$ 0.2   &  S,(2)  \\
& AC116  & 1984-12-21 & 9 & 2007+404 & 4.63  $\pm$ 0.01  &  4.9 &   9 & 27 & A   & 7.7   $\pm$ 0.2   &  S,(2) \\
& AF102  & 1985-05-09 & + & 2022+616 & 2.43  $\pm$ 0.01  & 20.4 &   9 & 26 & B   & 5.9   $\pm$ 0.2   &  S  \\
& AA47   & 1985-06-12 & 9 & 2007+404 & 3.87  $\pm$ 0.01  &  4.9 &   9 & 22 & B   & 5.6   $\pm$ 0.4   &  S,(2) \\
& AA47   & 1985-09-01 & 9 & 2007+404 & 3.53  $\pm$ 0.01  &  4.9 &   7 & 27 & C   &         $<$ 0.9   &  (2)  \\
& AA47   & 1985-09-19 & 9 & 2007+404 & 3.56  $\pm$ 0.01  &  4.9 &  11 & 25 & C   &         $<$ 0.6   &  (2)  \\
& VM115  & 1990-06-11 & 9 & 2202+422 & 3.24  $\pm$ 0.01  & 16.7 &  19 & 22 & A   &         $<$ 0.3   &  \\
& AG320  & 1991-03-06 & 9 & 2007+404 & 2.73  $\pm$ 0.01  &  5.2 &   7 & 25 & D   & 4.8   $\pm$ 0.5   &  \\
& AB671  & 1993-01-21 & 9 & 2007+404 & 3.09  $\pm$ 0.01  &  4.9 &  19 & 24 & A   & 2.3   $\pm$ 0.1   &  S \\
& AB671  & 1993-01-24 & 9 & 2007+404 & 3.08  $\pm$ 0.03  &  4.9 &  18 & 27 & A   & 2.1   $\pm$ 0.1   &  S \\
& AB671  & 1993-01-29 & 9 & 2007+404 & 3.12  $\pm$ 0.01  &  4.9 &  19 & 27 & BnA & 2.3   $\pm$ 0.1   &  S \\
& AB671  & 1993-02-01 & 9 & 2007+404 & 3.11  $\pm$ 0.01  &  4.9 &   7 & 27 & BnA & 2.2   $\pm$ 0.2   &  S \\
& AB671  & 1993-02-14 & 9 & 2007+404 & 3.12  $\pm$ 0.02  &  4.9 &   8 & 27 & BnA & 2.7   $\pm$ 0.2   &  S  \\
& AS483  & 1993-05-01 & 98& 2007+404 & 3.04  $\pm$ 0.01  &  4.9 &  21 & 25 & B   & 4.5   $\pm$ 0.4   &  (5) \\
& TST6CM & 1993-06-04 & 9 & 2052+365 & 5.2   $\pm$ 0.2   &  6.0 &  29 & 26 & CnB & 6.7   $\pm$ 0.4   &  X  \\
& AR277  & 1994-04-17 & 9 & 2007+404 & 3.04  $\pm$ 0.01  &  4.9 &  23 & 15 & A   & 8.0   $\pm$ 0.2   &  S,(4) \\
& AS544  & 1994-09-13 & 12& 2007+404 & 3.16  $\pm$ 0.01  &  4.8 &   5 & 25 & B   &        $<$ 10.    &  S \\
& AS544  & 1994-10-12 & 12& 2007+404 & 3.22  $\pm$ 0.01  &  4.8 &   5 & 25 & CnB & 3.9   $\pm$ 0.9   &  S  \\
& AR328  & 1995-04-27 & 9 & 2007+404 & 2.982 $\pm$ 0.003 &  4.9 &  11 & 16 & D   & 1.5   $\pm$ 0.4   &  (4) \\
& AS644  & 1999-02-05 & 7 & 2007+404 & 2.6   $\pm$ 0.1   &  4.9 &  20 & 23 & DnC & 5.3   $\pm$ 1.4   &  \\
& AS644  & 1999-04-18 & 9 & 2015+371 & 1.728 $\pm$ 0.003 &  5.3 &   5 & 27 & D   & 7.1   $\pm$ 0.5   &  S \\
& AW515  & 1999-06-08 & 9 & 2007+404 & 2.40  $\pm$ 0.03  &  4.9 &  10 & 21 & AD  & 5.3   $\pm$ 0.6   &  I,S  \\
& AC530  & 1999-09-03 & 12& 2007+404 & 2.53  $\pm$ 0.05  &  4.8 &  49 &  9 & A   &        $<$ 11.    &  S \\
& AU082  & 2000-06-30 & 12& 2015+371 & 1.92  $\pm$ 0.02  &  5.2 &  88 &  7 & DnC & 6.    $\pm$ 4.    &  \\
& AS786  & 2004-02-15 & 7 & 2015+371 & 2.78  $\pm$ 0.003 &  5.4 &  61 & 26 & CnB & 7.    $\pm$ 2.    &  \\
& AB1156 & 2005-02-04 & 8 & 2007+404 & 2.47  $\pm$ 0.01  &  4.9 &  31 & 25 & BnA & 4.7   $\pm$ 0.3   &  \\
& AB1156 & 2005-02-06 & 8 & 2007+404 & 2.50  $\pm$ 0.02  &  4.9 &  31 & 25 & BnA & 4.4   $\pm$ 0.6   &  \\
& AB1156 & 2005-02-08 & 8 & 2007+404 & 2.52  $\pm$ 0.01  &  4.9 &  31 & 25 & BnA & 4.1   $\pm$ 1.1   &  \\
& AB1156 & 2005-02-11 & 8 & 2007+404 & 2.52  $\pm$ 0.01  &  4.9 &  31 & 25 & BnA & 5.0   $\pm$ 0.9   &  \\
& AB1156 & 2005-02-13 & 8 & 2007+404 & 2.49  $\pm$ 0.02  &  4.9 &  31 & 24 & BnA & 4.3   $\pm$ 0.3   &  \\
& AB1156 & 2005-02-18 & 8 & 2007+404 & 2.44  $\pm$ 0.01  &  4.9 &  51 & 24 & B   & 4.1   $\pm$ 0.2   &  \\
& AB1156 & 2005-02-25 & 8 & 2007+404 & 2.49  $\pm$ 0.01  &  4.9 &  30 & 24 & B   & 5.5   $\pm$ 0.3   &  \\
& AB1156 & 2005-02-27 & 8 & 2007+404 & 2.49  $\pm$ 0.01  &  4.9 &  51 & 25 & B   & 4.6   $\pm$ 0.2   &  \\
& AB1156 & 2005-02-28 & 8 & 2007+404 & 2.47  $\pm$ 0.01  &  4.9 &  30 & 25 & B   & 4.5   $\pm$ 0.2   &  \\
& AB1156 & 2005-03-01 & 8 & 2007+404 & 2.410 $\pm$ 0.004 &  4.9 &  31 & 24 & B   & 4.4   $\pm$ 0.3   &  \\
& AB1156 & 2005-03-02 & 8 & 2007+404 & 2.44  $\pm$ 0.01  &  4.9 &  29 & 25 & B   & 4.4   $\pm$ 0.2   &  \\
& AB1156 & 2005-03-08 & 8 & 2007+404 & 2.44  $\pm$ 0.01  &  4.9 &  26 & 25 & B   & 4.7   $\pm$ 0.2   &  \\
& AB1156 & 2005-03-12 & 8 & 2007+404 & 2.43  $\pm$ 0.01  &  4.9 &  30 & 25 & B   & 5.1   $\pm$ 0.5   &    \\
& BD110  & 2005-11-15 & 9 & 2007+404 & 2.51  $\pm$ 0.01  &  4.9 &  18 & 21 & D   & 6.8   $\pm$ 0.4   &  S  \\
\multicolumn{5}{l}{{\bf 20 cm}}\\*
& VAND   & 1981-06-22 & A & 2022+616 & 2.18  $\pm$ 0.01  & 20.2 & 6   & 26 & B   &         $<$ 9.    &  1x3.125,S \\
& BECK   & 1982-07-03 & 9 & 2007+404 & 3.99  $\pm$ 0.02  &  4.9 & 4   & 27 & A   & 5.2   $\pm$ 0.7   &  1x50,(1) \\
& BIEG   & 1982-08-26 & 9 & 2007+404 & 4.00  $\pm$ 0.04  &  4.9 & 34  & 24 & B   & 4.4   $\pm$ 0.3   &  1x50,(2) \\
& BECK   & 1982-12-21 & 9 & 2023+318 & 2.18  $\pm$ 0.02  &  9.6 & 15  & 26 & D   &        $<$ 24.    &  1x50,S \\
& AC42   & 1983-05-09 &12 & 2007+404 & 4.33  $\pm$ 0.03  &  4.8 & 1   & 27 & C   &        $<$  6.    &  1x50,(2) \\
& AB228  & 1983-08-25 & 9 & 2007+404 & 4.19  $\pm$ 0.02  &  4.9 & 28  & 24 & A   & 0.5   $\pm$ 0.2   &  \\
& AB228  & 1983-08-27 & 9 & 2007+404 & 4.31  $\pm$ 0.02  &  4.9 & 28  & 26 & A   & 0.5   $\pm$ 0.2   &  \\
& AR96O  & 1984-01-06 & 5 & 2007+404 & 3.39  $\pm$ 0.04  &  4.7 & 27  & 15 & B   &        $<$  7.    &  \\
& AA28   & 1984-03-09 & 8 & 2007+404 & 4.15  $\pm$ 0.02  &  4.9 & 26  & 26 & CnB & 5.    $\pm$ 3.    &  \\
& AM119  & 1984-06-11 & X3& 2007+404 & 4.2   $\pm$ 0.1   &  4.7 & 20  & 27 & C   & 5.8   $\pm$ 1.0   &  2x25,S \\
& AM119  & 1984-06-12 & X3& 2007+404 & 4.40  $\pm$ 0.03  &  4.7 & 21  & 27 & C   & 3.9   $\pm$ 0.5   &  2x25,S \\
& AR110  & 1984-09-06 & 5 & 2007+404 & 3.99  $\pm$ 0.03  &  4.7 & 13  & 25 & D   &        $<$ 10.    &  S \\
& AR110  & 1984-09-15 & 5 & 2007+404 & 3.86  $\pm$ 0.08  &  4.7 & 34  & 26 & D   &        $<$ 11.    &  S \\
& AR110  & 1984-09-20 & 5 & 2007+404 & 4.2   $\pm$ 0.1   &  4.7 & 16  & 27 & D   &        $<$ 20.    &  S \\
& AR110  & 1984-09-22 & 5 & 2007+404 & 4.00  $\pm$ 0.06  &  4.7 & 17  & 26 & D   &        $<$ 13.    &  S \\
& AR110  & 1984-09-24 & 5 & 2007+404 & 4.06  $\pm$ 0.04  &  4.7 & 45  & 26 & D   &        $<$ 12.    &  S \\
& AR110  & 1984-09-28 & 5 & 2007+404 & 4.04  $\pm$ 0.03  &  4.7 & 41  & 26 & D   &        $<$ 10.    &  S \\
& AC116  & 1984-11-27 & 9 & 2007+404 & 4.06  $\pm$ 0.02  &  4.9 & 33  & 25 & A   & 6.2   $\pm$ 0.2   &  (2) \\
& AC116  & 1984-12-21 & 9 & 2007+404 & 3.98  $\pm$ 0.03  &  4.9 & 13  & 27 & A   & 4.9   $\pm$ 0.2   &  (2) \\
& AT61   & 1985-04-04 & 9 & 2052+365 & 5.43  $\pm$ 0.01  &  6.0 & 14  & 25 & BnA & 4.3   $\pm$ 0.2   &  \\
& AA47   & 1985-09-19 & 9 & 2007+404 & 3.92  $\pm$ 0.04  &  4.9 & 13  & 26 & C   &        $<$  4.5   &  S,(2) \\
& AR170  & 1987-10-16 & 5 & 2052+365 & 5.20  $\pm$ 0.02  &  6.2 & 26  & 23 & BnA & 2.2   $\pm$ 1.0   &  \\
& AR170  & 1987-11-09 & 5 & 2052+365 & 5.07  $\pm$ 0.05  &  6.2 & 9   & 21 & BnA & 6.    $\pm$ 4.    &  \\
& AR170  & 1988-02-05 & 5 & 2052+365 & 4.38  $\pm$ 0.07  &  6.2 & 16  & 16 & B   &        $<$  9.    &  \\
& ADHOC  & 1988-05-03 & 23& 2007+404 & 3.37  $\pm$ 0.01  &  4.7 & 3   & 27 & C   &        $<$ 16.    &  2x25,S \\
& VP51   & 1989-06-19 & 9 & 2052+365 & 5.71              &  6.0 & 26  & 26 & C   & 5.3   $\pm$ 0.7   &  PH,S \\
& VP91   & 1989-11-10 & 9 & 2007+404 & 3.3   $\pm$ 0.1   &  4.9 & 30  & 27 & D   &        $<$ 18.    &  S,(3) \\
& AM305  & 1990-07-05 & 80& 2007+404 & 2.84  $\pm$ 0.02  &  4.9 & 5   & 26 & BnA &        $<$  3.4   &  2x25 \\
& AT107  & 1990-08-09 & 21& 2052+365 & 5.15  $\pm$ 0.01  &  6.0 & 4   & 27 & B   &        $<$ 30.    &  \\
& AK279  & 1991-03-29 & 21& 2052+365 & 7.8   $\pm$ 0.3   &  6.0 & 31  & 26 & D   &        $<$ 30.    &  S \\
& AK279  & 1991-03-31 & 21& 2052+365 & 6.5   $\pm$ 0.2   &  6.0 & 21  & 26 & D   &        $<$ 37.    &  S \\
& AFTST  & 1991-08-11 & 21& 2052+365 & 5.15  $\pm$ 0.01  &  6.0 & 38  & 19 & A   &        $<$  9.    &  \\
& AW311  & 1992-03-15 & 21& 2022+542 & 1.036 $\pm$ 0.002 & 13.2 & 29  & 27 & C   & 3.7   $\pm$ 1.4   &  S \\
& BP1    & 1992-05-30 & 9 & 2052+365 & 5.1   $\pm$ 0.3   &  6.0 & 129 & 25 & DnC &        $<$ 11.    &  \\
& AS483  & 1993-04-29 & X3& 2007+404 & 2.52  $\pm$ 0.02  &  4.7 & 14  & 25 & B   &        $<$ 16.    &  S \\
& TST6CM & 1993-06-04 & 9 & 2052+365 & 6.4   $\pm$ 0.3   &  6.0 & 13  & 27 & CnB &        $<$  5.    &  S \\
& AL372  & 1996-03-28 & 5 & 2038+513 & 5.80              & 10.1 & 10  & 25 & C   & 1.9   $\pm$ 1.0   &  2x3.125,PH,S  \\
& AC308  & 1996-09-15 & 91& 2202+422 & 6.15  $\pm$ 0.03  & 16.9 & 1   & 27 & D   &        $<$150.    &  \\
& AC308  & 1996-09-20 & 32& 1924+334 & 3.793 $\pm$ 0.003 & 15.6 & 1   & 27 & D   &        $<$ 19.    &  \\
& AC496  & 1997-09-27 & 20& 1924+334 & 3.753 $\pm$ 0.004 & 15.6 & 3   & 26 & DnC &        $<$ 16.    &  \\
& AA237  & 1999-02-01 & J2& 2052+365 & 5.31  $\pm$ 0.03  &  6.1 & 11  & 27 & C   & 9.    $\pm$ 2.    &  \\
& AR458  & 2001-02-06 & X3& 2015+371 & 1.56  $\pm$ 0.01  &  6.0 & 32  & 26 & BnA &        $<$ 10.    &  S \\
& AB1075 & 2003-04-29 & SE& 2052+365 & 4.98  $\pm$ 0.01  &  6.1 & 30  & 27 & D   &        $<$ 16.    &  S \\
& BD110  & 2005-11-15 & 9 & 2007+404 & 2.6   $\pm$ 0.1   &  4.9 & 20  & 22 & D   &        $<$ 24.    &  S \\
\multicolumn{5}{l}{{\bf 90 cm}}\\*
& SYSTE  & 1986-05-20 &X3 & 2052+365 & 2.18  $\pm$ 0.02  &  5.9 & 158 & 12 & A   &        $<$  8.    &  2x3.125 \\
\hline
\end{longtable}
}

\end{document}